\documentstyle{article}\begin{document}\centerline{\bf A note on De Moivre's quintic equation}\vskip .4in

\centerline{M. L. Glasser}
\vskip .1in

\centerline{Department of Physics, Clarkson University}                                                                                                                                                                                                                 
,
\centerline{Potsdam, NY 13699-5820 (USA)}\vskip .3in

\begin {quote} The quintic equation with real coefficients
$$x^5+5ax^3+5a^2x+b=0$$
is solved in radicals and the result is used to sum a hypergeometric series for several arguments..\end{quote}

\newpage

 We start with the general quintic equation (over the real numbers)
 $$\sum_{j=0}^5a_j x^j=0\eqno(1)$$
 with $a_5=1$. Using Viet\'a's substitution, $x=u-a/u$ with $a\ne0$ real, expanding and multiplying by $u^5$, the resulting 10-th degree equation is solvable (in radicals) only for $a_1=aa_3=5a^2$, $a_2=a_4=0$,
 in which case
 $$u^{10}+a_0u^5-a^5=0.\eqno(2)$$ By solving this quadratic equation (in $u^5$) and extracting the fifth roots, after eliminating duplicates, we see that the roots of the De Moivre equation (writing $b$ in place of $a_0$)
 $$x^5+5ax^3+5a^2x+b=0\eqno(3)$$
 are
 $$x_1=2^{-1/5}[(\sqrt{\gamma}-b)^{1/5}-(\sqrt{\gamma}+b)^{1/5}]$$
 $$x_2=-2^{-1/5}[e^{\pi i/5}(\sqrt{\gamma}-b)^{1/5}+e^{4\pi i/5}(\sqrt{\gamma}+b)^{1/5}]=x_4^*$$
 $$x_3=2^{-1/5}[e^{2\pi i/5}(\sqrt{\gamma}-b)^{1/5}+e^{3\pi i/5}(\sqrt{\gamma}+b)^{1/5}]=x_5^*.$$
 $$\gamma=4a^5+b^2.\eqno(4)$$
 (The asterisk denotes the complex conjugate.)
 
 By means of the Tschirnhausen transformation [1] $y=x^2+\alpha x+\beta$ any quintic can be reduced to ``principal" form $y^5+Ay^2+By+C=0$. This can be carried out efficiently using Mathematica by means of the command 
 
 $Resultant[x^5+5ax^3+5a^2x+b,y-(x^2+\alpha x+\beta),x]$
 which returns $y^5+(10a-5\beta)y^4+(35a^2+5a\alpha^2-40a\beta+10\beta^2)y^3+\dots$ from which we easily determine $\alpha=\sqrt{a}$ and $\beta=2a$. Thus, the principal form for the De Moivre quintic is
 $$y^5+(5\sqrt{a}b-10a^3)y^2+15a^4y-(9a^{5/2}b+22a^5+b^2)=0.\eqno(5)$$
 
 Jerrard[1] found that by means of a quartic Tschirnhausen transformation a principal quintic can be further reduced to the trinomial Bring form
 $$z^5-z+t=0.\eqno(6)$$
 The details can be mechanized to some extent[2], but  are still quite cumbersome. To present the result it is convenient to introduce the further abbreviations:
 $\alpha=8a^{5/2}-b$, $\beta=2a^{5/2}-b$, $\delta=176a^5+36a^{5/2}b-b^2$.  Also $d_0=675(\beta/\alpha)(2a^{5/2}+b)a^2b $, $d_1=\Delta/\alpha$,  $d_2=25(\beta/\alpha)(8a^5-3a^{5/2}b-b^2)$, $d_3=75(\beta/\alpha)a^{3/2}(16a^{5/2}+3b)$ and $d_4=75\beta a^{1/2}$, where
 $$\Delta=25[\frac{\delta\gamma^{1/2}\alpha}{3^{2/3}a^{1/2}}(\beta\sqrt{3a^{3/2}(11a^{5/2}-b)}-9a^2\gamma^{1/2})^{1/3}+$$
 $$\frac{1}{3^{1/3}\alpha\gamma^{1/2}}\frac{16a^{15/2}+4a^5b+4a^{5/2}b^2+b^3}{(\beta\sqrt{3a^{3/2}(11a^{5/2}-b)}-9a^2\gamma^{1/2})^{1/3}}-$$
 $$(800a^{17/2}-318a^6b+227a^{7/2}b^2-12ab^3)].\eqno(7)$$
 The quantity $\Delta$ has six possible values, due to the choice of branches in the square and cube roots, so it takes a bit of experimentation to select the appropriate one. 
 
 Next, we require the coefficients
 $$c_0=\frac{5^6\delta^3\gamma^2}{\alpha^5}[5625\beta^2f_1(a,b)+\frac{25}{a}\beta\Delta f_2(a,b)+3a^{1/2}\Delta^2(188a^5+86a^{5/2}b+9b^2)]\eqno(8a)$$
 $$c_1=\frac{5^5\delta^2\gamma a^{1/2}}{\alpha^4}[5625\beta^2g_1(a,b)+\frac{25}{a}\beta\Delta g_2(a,b)+9a^{3}\Delta^2(4a^{5/2}+b)],\eqno(8b)$$

 where
 
 $$f_1(a,b)=2622464a^{35/2}-1339776a^{15}b+103828a^{25/2}b^2+218210a^{10}b^3-$$
 $$17365a^{15/2}b^4-2858a^5b^5+297a^{5/2}b^6-b^7$$
 $$f_2(a,b)=209024a^{25/2}-30616a^{10}b-34508a^{15/2}b^2-$$
 $$98a^5b^3+530a^{5/2}b^4-b^5,$$
 $$g_1(a,b)=133120a^{35/2}-83520a^{15}b+26804a^{25/2}b^2+1789a^{10}b^2-$$
 $$1082a^{15/2}b^4+386a^5b^5-48a^{5/2}b^6+b^7,\eqno(9)$$
 $$g_2(a,b)=11200a^{25/2}-3656a^{10}b-754a^{15/2}b^2+62a^5b^3-38a^{5/2}b^4+b^5.$$
 \vskip .1in
 
 The result is that the roots of the Bring-Jerrard quintic
 $$z^5-z+t=0\eqno(10)$$
 with
 $$t=-e^{-i\pi/4}c_0c_1^{-5/4}\eqno(11)$$
 are
 $$z_j=e^{-i\pi/4}c_1^{-1/4}\sum_{k=0}^4d_k(x_j^2+\sqrt{a}x_j+2a)^k.\eqno(12).$$
 
 \vskip .1in
     Now it has been shown by James Cockle[3] ( the first Chief Magistrate of Queensland) and others[4]
  that one of the roots of (10) is
  $$z_0=   t\;_4F_3(1/5,2/5,3/5,4/5;1/2,3/4,5/4;\frac{3125}{256}t^4).\eqno(13)$$
 Therefore,  it has been shown that
 $$\;_4F_3(1/5,2/5,3/5,4/5;1/2,3/4,5/4;-\frac{3125}{256}\frac{c_0^4}{c_1^5})=$$
 $$-\frac{c_1}{c_0}\sum_{k=0}^4d_k(x_0^2+\sqrt{a}x_0+2a)^k\eqno(14)$$
 where $x_0$ is one of the five values in (4). (Which one it is can be determined numerically).
 \vskip .1in
 
 As a very simple example, let us take the case  $b=2a^{5/2}$. Then the principal quintic (5) is already in Bring form, which is easily scaled by $y=e^{i\pi/4}15^{1/4}az$ into (6) with $t=44e^{-i\pi/4}15^{-5/4}$. Therefore, from (13) we find
 $$\;_4F_3(1/5,2/5,3/5,4/5;1/2,3/4,5/4;-\frac{14641}{243})=$$
 $$\frac{15}{44}[(\sqrt{2}+1)^{2/5}+(\sqrt{2}-1)^{2/5}-(\sqrt{2}+1)^{1/5}+(\sqrt{2}-1)^{1/5}].\eqno(15)$$
 
 At the other extreme, the algorithm described here [5]  produces formulas such as

$$\;_4F_3(1/5,2/5,3/5,4/5;1/2,3/4,5/4;z)=Z$$
where

$$z=((13 + \sqrt{182})^{2/3}
    ((789 13^{2/3} + 247 13^{1/6} \sqrt{14}) (13 + \sqrt{182})^{1/3}
       + 2275 (13 + \sqrt{182})^{2/3} - $$
    $$ 13^{1/3} (6799 + 542 \sqrt{182}))^4)/(1521 ((-7 13^{2/3} + 
        3 13^{1/6} \sqrt{14}) (13 + \sqrt{182})^{1/3} - $$
   $$  161 (13 + \sqrt{182})^{2/3} + 13^{1/3} (133 + 10 \sqrt{182}))^5)$$
   and
 $$ Z=(5 ((-7 13^{2/3} + 3 13^{1/6} \sqrt{14}) (13 + \sqrt{182})^{1/3} - $$
    $$ 161 (13 + \sqrt{182})^{2/3} + $$
     $$13^{1/3} (133 + 10 \sqrt{182})) (-3 (-5 + \sqrt{26})^{4/5}
       (13 + \sqrt{182})^{1/3}
       (1103 + 3465 (5 + \sqrt{26})^{1/5} + $$
     $$   5355 (5 + \sqrt{26})^{2/5} + 3780 (5 + \sqrt{26})^{3/5}) + $$
   $$  3 (-5 + \sqrt{26})^{3/5} (13 + \sqrt{182})^{1/3}
       (8872 + 1485 \sqrt{26} + 4412 (5 + \sqrt{26})^{1/5} + $$
      $$  6930 (5 + \sqrt{26})^{2/5} + 3780 (5 + \sqrt{26})^{4/5}) + (-5 +
         \sqrt{26})^{1/5}
       (-142 13^{2/3} + $$
      $$  142 13^{1/3} (13 + \sqrt{182})^{2/3} + (13 + \sqrt{182})^{1/3}
          (34114 + 5355 \sqrt{26} + $$
         $$13236 (5 + \sqrt{26})^(3/5) + 
           10395 (5 + \sqrt{26})^{4/5} - $$
         $$  9 (5 + \sqrt{26})^{2/5} (-2797 + 180 \sqrt{26}))) - (-5 + 
        \sqrt{26})^{2/5}
       (-142 13^{2/3} + $$
       $$ 142 13^{1/3} (13 + \sqrt{182})^{2/3} + (13 + \sqrt{182})^{1/3}
          (33742 + 6480 \sqrt{26} +$$
         $$ 20790 (5 + \sqrt{26})^{3/5} + 
           16065 (5 + \sqrt{26})^{4/5} + $$
        $$  9 (5 + \sqrt{26})^{1/5} (2797 + 180 \sqrt{26}))) + (5 + \sqrt{
        26})^{1/5} (142 13^{2/3} - 
        3309 (5 + \sqrt{26})^{3/5} (13 + \sqrt{182})^{1/3} + $$
      $$  3 (5 + \sqrt{26})^{2/5}
          (-8872 + 1485 \sqrt{26}) (13 + \sqrt{182})^{1/3}
          +$$
         $$ (-34114 + 5355 \sqrt{26}) (13 + \sqrt{182})^{1/3} - 
        142 13^{1/3} (13 + \sqrt{182})^{2/3} + $$
       $$ 2 (5 + \sqrt{26})^{1/5}
          (71 13^{2/3}
            + (-16871 + 3240 \sqrt{26}) (13 + \sqrt{182})^{1/3} - 
           71 13^{1/3} (13 +$$
           $$ \sqrt{182})^{2/3}))))/(568 (13 + \sqrt{
     182})^{1/3}
     ((789 13^{2/3} +$$
    $$ 247 13^{1/6} \sqrt{14}) (13 + \sqrt{182})^{1/3}
       + 2275 (13 + \sqrt{182})^{2/3} - 
     13^{1/3} (6799 + 542 \sqrt{182})))$$
     
    \noindent
 {\bf Acknowledgement} I am grateful to Prof. Victor Adamchik for help with the necessary Mathematica routines.
 \newpage
 \centerline{\bf References}\vskip .1in
 
 \noindent
 [1] V.S. Adamchik and D.J. Jeffrey,  ACM SIGSAM Bulletin{\bf{37}}, 90 (2003).
 
 \noindent
 [2] Victor Adamchik, Private communication.
 
 \noindent
 [3] J. Cockle, Phil. Mag.{\bf{20}}, 145 (1860).
 
 \noindent
 [4] M.L. Glasser, J. Comp. Appl. Math., {\bf 118}, 169 (2000).
 
 \noindent
 [5] The author has put together a Mathematica notebook incorporating this procedure which will be provided on request to laryg@clarkson.edu.

 \end{document}